\documentclass[10pt,fleqn]{article}
\pdfoutput=1
\usepackage{psfrag}
\usepackage{amsmath}
\usepackage[dvips]{epsfig}
\usepackage{amsmath,amssymb,bm}
\usepackage{graphics}
\usepackage[pdftex,colorlinks,bookmarks,pdfpagemode=UseOutlines,linkcolor=black,
urlcolor=black,citecolor=black,pageanchor=false]{hyperref}
\usepackage[margin=1in]{geometry}
\usepackage[T1]{fontenc}
\usepackage{fix-cm}
\usepackage[authoryear,round]{natbib}
\usepackage[pagewise,mathlines]{lineno}
\usepackage{amsthm}
\usepackage{longtable}
\usepackage{xfrac}
\usepackage{booktabs}
\usepackage{multirow}

\usepackage{placeins}
\let\Oldsection\section
\renewcommand{\section}{\FloatBarrier\Oldsection}

\let\Oldsubsection\subsection
\renewcommand{\subsection}{\FloatBarrier\Oldsubsection}

\newtheorem{theo}{Theorem}[section]
\newtheorem{lemma}{Lemma}[section]
\newtheorem{df}{Definition}[section]
\newtheorem{cor}{Corollary}[section]
\newtheorem{assump}{Assumption}[section]
\newtheorem{assert}{Assertion}[section]
\newtheorem{remark}{Remark}[section]
\newcommand{\bl}{\begin{lemma}}
\newcommand{\el}{\end{lemma}}
\newcommand{\be}{\begin{equation}}
\newcommand{\ee}{\end{equation}}
\newcommand{\beqn}{\begin{eqnarray}}
\newcommand{\eeqn}{\end{eqnarray}}
\newcommand{\bt}{\begin{theo}}
\newcommand{\et}{\end{theo}}
\newcommand{\bd}{\begin{df}}
\newcommand{\ed}{\end{df}}
\newcommand{\ba}{\begin{assump}}
\newcommand{\ea}{\end{assump}}
\newcommand{\bass}{\begin{assert}}
\newcommand{\eass}{\end{assert}}
\newcommand{\brem}{\begin{remark}}
\newcommand{\erem}{\end{remark}}
\newcommand{\bc}{\begin{cor}}
\newcommand{\ec}{\end{cor}}

\newcommand{\BB}{{\cal B}}

\newcommand{\pt}{\tilde{p}}

\newcommand{\St}{\tilde{S}}
\newcommand{\So}{\overline{S}}

\newcommand{\Ut}{\tilde{U}}
\newcommand{\Vt}{\tilde{V}}
\newcommand{\OBB}{\overline{\BB}}

\newcommand{\etab}{\bm{\eta}}

\numberwithin{equation}{section}

\long\def\comment#1{}

\nolinenumbers

\title{Macroscopic Modeling, Calibration, and Simulation of Managed Lane-Freeway Networks, Part I: Topological and Phenomenological Modeling
\author{Matthew A. Wright, Roberto Horowitz, Alex A. Kurzhanskiy}
\date{}
}

\begin{document}
\maketitle

\begin{abstract}
To help mitigate road congestion caused by the unrelenting growth of traffic demand, many transit authorities have implemented managed lane policies. Managed lanes typically run parallel to a freeway's standard, general-purpose (GP) lanes, but are restricted to certain types of vehicles. It was originally thought that managed lanes would improve the use of existing infrastructure through incentivization of demand-management behaviors like carpooling, but implementations have often been characterized by unpredicted phenomena that is often to detrimental system performance. Development of traffic models that can capture these sorts of behaviors is a key step for helping managed lanes deliver on their promised gains. 

Towards this goal, this paper presents several macroscopic traffic modeling tools we have used for study of freeways equipped with managed lanes, or ``managed lane-freeway networks.'' The proposed framework is based on the widely-used first-order kinematic wave theory. In this model, the GP and the managed lanes are modeled as parallel links connected by nodes, where certain type of traffic may switch between GP and managed lane links. Two types of managed lane configuration are considered: full-access, where vehicles can switch between the GP and the managed lanes anywhere; and \emph{separated}, where such switching is allowed only at certain locations called \emph{gates}. We describe macroscopic modeling considerations for both types of network topologies.

We also describe methods to incorporate three phenomena into our model that are particular to managed lane-freeway networks: the \emph{inertia effect}, the \emph{friction effect}, and the \emph{smoothing effect}. The inertia effect reflects drivers' inclination to stay in their lane as long as possible and switch only if this would obviously improve their travel condition. The friction effect reflects the empirically-observed driver fear of moving fast in a managed lane while traffic in the adjacent GP lanes moves slowly due to congestion. Finally, the smoothing effect describes how managed lanes can \emph{increase} throughput at bottlenecks by reducing lane changes. We present simple modeling techniques for each of these phenomena that fit within the general macroscopic theory for arbitrary link and junction models.

In this paper's sequel, we present model calibration methodologies and simulation results.

\end{abstract}


\section{Introduction}\label{sec_intro}
Traffic demand in the developed and developing worlds shows no sign of decreasing, and the resulting congestion remains a costly source of inefficiency in the built environment.
One study \citep{schrank_2015_2015} estimated that, in 2014, delays due to congestion cost drivers 7 billion hours and \$160B in the United States alone, leading to the burning of 3 billion extra gallons of fuel.
The historical strategy for accommodating more demand has been construction of additional infrastructure, but in recent years planners have also developed strategies to improve the performance of \emph{existing} infrastructure, both through improved road operations and \emph{demand management}, which seeks to lower the number of vehicles on the road \citep{kurzhanskiy_traffic_2015}.
One such strategy that has been widely adopted in the United States and other developed countries is the creation of so-called \emph{managed lanes} \citep{obenberger_managed_2004}.
Managed lanes are implemented on freeways by restricting the use of one or more lanes to certain vehicles.
As an example, high-occupancy-vehicle (HOV) lanes are intended to incentivize carpooling, which reduces the total number of cars on the road as a demand management outcome \citep{chang_review_2008}.

In addition to demand management, managed lanes provide an opportunity for improved road operations through real-time, responsive traffic control.
For example, tolled express lanes give drivers the opportunity to pay a toll to drive parallel to the general-purpose lanes on a (presumably less-congested) express lane.
Traffic management authorities here have an opportunity to adjust the toll amount in response to the real-time state of traffic on the network.
The potential for managed lanes as instruments for reactive, real-time traffic operations--in addition to their demand-management purpose--has made them popular among transportation authorities \citep{kurzhanskiy_traffic_2015}.

However, the traffic-operational effects of managed lanes are not always straightforward or as rehabilitative as expected, as their presence can create complex traffic dynamics.
Even in a freeway with simple geometry, the dynamics of traffic flow are complex and not fully understood, and adding managed lanes alongside the non-managed, general-purpose (GP) lanes only exacerbates this.
In effect, adding a managed lane creates two parallel and distinct, but coupled, traffic flows on the same physical structure.
When used as intended, managed lanes carry flows with different density-velocity characteristics and vehicle-type (e.g., strictly HOVs) compositions than the freeway.
When vehicles move between the two lane flows, these two heterogeneous flows mix, and complex phenomena that are unobserved in GP-only freeways can emerge (see, e.g., \citet{menendez_effects_2007, daganzo_effects_2008, cassidy_smoothing_2010, cassidy_problem_2015, liu_analysis_2011, jang_dual_2012, thomson_operational_2012,ponnu_speed-spacing_2015,ponnu_when_2017,fitzpatrick_operating_2017,kim_safety_2018}, and others).

Making better use of managed lanes requires an understanding of the macroscopic behavior they induce.
One widely-used tool for understanding macroscopic traffic flow behavior is the macroscopic traffic flow model.
A rich literature exists on macroscopic models for flows on long roads, and at junctions where those roads meet, but an extension to the parallel-flows situation created by placing a managed lane in parallel with a freeway (a ``managed lane-freeway network'') is not straightforward.
For example, the GP/managed lane interface has been observed to exhibit unique phenomena, such as e.g., 1) a ``friction effect'' \citep{liu_analysis_2011,jang_traffic_2012}, where vehicles in a GP lane adjacent to the managed lane(s) move slower than what would be expected based solely on their density of cars \citep{liu_analysis_2011,fitzpatrick_operating_2017} (a common theory is that these GP lane vehicles move slowly out of fear that vehicles may suddenly move out of the managed lane in front of them), or 2) a ``smoothing effect'' at bottlenecks \citep{menendez_effects_2007, cassidy_smoothing_2010,jang_dual_2012} that leads to an \emph{increased} flow in the GP lanes closest to the managed lane(s) by reducing lane changes (briefly, fewer vehicles being eligible for the fastest lane means that fewer vehicles will change lanes - and in the process of changing lanes, slow down surrounding traffic - to enter it. See, e.g., \citet{zheng_recent_2014} for more information on the effects of lane changes on macroscopic flow characteristics).
In the present paper, we propose simple models for these emergent phenomena that fit within the classic macroscopic kinematic wave theory.
To the best of our knowledge, we present the first macroscopic modeling technique for the smoothing effect, and the first proposed model of the friction effect that considers junctions between the managed and GP lanes (we will discuss the differences between our proposed friction effect model and ones previously proposed in the literature below).

This paper presents macroscopic flow modeling tools we have used for simulation of managed lane-freeway networks.
We begin in Section \ref{sec_lit_review} with a discussion of relevant modeling tools from the literature, and how we make use of them.
Section \ref{sec_hov_model} describes network structures for the two common managed lane configurations: 
gated-access and full-access (ungated) lanes.
Section \ref{sec_phenomena} describes simple models for the emergent phenomena that are particular to managed lane-freeway networks: the friction, smoothing, and inertia effects.
In Part II of this paper series \citep{wright_macroscopic_2019_pt2}, we put all of these pieces together, discuss how to calibrate the full model, and present modeling case studies of two freeways with managed lanes in California.

\section{Managed Lane Modeling}\label{sec_lit_review}
The modeling techniques presented in this paper are based on the first-order ``kinematic wave'' macroscopic traffic flow model.
These models describe aggregate traffic flows as fluids following a one-dimensional conservation law.
We briefly introduce our notation here, but do not discuss the basics of this class of models.
Detailed reviews are available in many references.

\subsection{Modeling basics}\label{subsec_modelingbasics}
In this simulation framework, a road is divided into discrete cells, which we refer to as \emph{links}.
Links are drawn between \emph{nodes}: a link begins at one node and ends at another.
Many links may begin and end at each node.
Each link $l$ is characterized by density $n_l$, the number of cars in the link.
In a first-order model, the traffic flows are fully prescribed by the density.
From timestep $t$ to $t+1$, link $l$'s density updates according to the equation
\begin{linenomath}
\begin{equation}
	n_l(t+1) = n_l(t) + \frac{1}{L_l} \left( \sum_{i=1}^M f_{il}(t) - \sum_{j=1}^N f_{lj}(t) \right), \label{eq:ctmbasic}
\end{equation}
\end{linenomath}
where $L_l$ is the length of link $l$, $f_{il}(t)$ is the flow (number of vehicles) leaving link $i$ and entering link $l$ at time $t$, $f_{lj}(t)$ is the flow leaving link $l$ and entering link $j$ at time $t$, $M$ is the number of links that end at link $l$'s beginning node, and $N$ is the number of links that begin at link $l$'s ending node.

Computing the inter-link flows requires the use of two intermediate quantities for each link.
These are the link demand $S_l(t)$, which is the number of vehicles that wish to exit link $l$ at timestep $t$; and the link supply, $R_l(t)$, which is the number of vehicles link $l$ can accept at time $t$.
Both $S_l$ and $R_l$ are functions of the density $n_l$.
The model that computes $S_l$ and $R_l$ from $n_l$ is often called the ``fundamental diagram'' or ``link model,'' and the model that computes the flows from all links' supplies and demands is often called the ``node model.''

A brief outline of how first-order macroscopic simulation of a road network (sometimes called a \emph{dynamic network loading} simulation) is performed could be:
\begin{enumerate}
  \item At time $t$, use the link model for each link $l$ to compute the link's demand $S_l(t)$ and supply $R_l(t)$ as a function of its density $n_l(t)$.
  \item Use the node model for each node to compute the inter-link flows $f_{ij}$ for all incoming links $i$ and outgoing links $j$ as functions of $S_i(t)$, $R_j(t)$, and information about vehicles' desired movements $ij$.
  \item Update the state of each link using~\eqref{eq:ctmbasic}.
  \item Increment $t$ and repeat until the desired simulation end time is reached.
\end{enumerate} 

The tools described in this paper are compatible with any such link model.
We make use of a particular node model that we have studied in \citet{wright_dynamic_2016,wright_node_2017}.
An aspect of this node model of particular relevance to managed lane modeling is our ``relaxed first-in-first-out (FIFO) rule'' construction~\citep{wright_node_2017}.
This is necessary for modeling the flows between the GP and managed lanes.
Without a FIFO relaxation, congestion in one of the two lane groups could block traffic in the other when that may be unrealistic (see~\citet[Section 2.2]{wright_node_2017} for a detailed discussion).

So far, we have presented ingredients for a model that, while able to express many simple network topologies by joining links and nodes, does not capture several important behaviors in managed lane-freeway networks.
The next three Sections briefly overview the additions to the standard model that will be explained in greater detail in the remainder of the paper.

\subsection{Multiple classes of vehicles and drivers}\label{subsec_multipleclasses}
In \eqref{eq:ctmbasic}, we describe the number of vehicles in a link as a single number, $n_l$.
In this formulation, all vehicles are treated the same.
However, for simulation in a managed lane-freeway network, it makes sense to break $n_l$ into different classes of vehicles and/or drivers.
For example, for a freeway with an HOV lane facility, we might consider two classes: HOVs and non-HOVs.
To this end, \eqref{eq:ctmbasic} can be rewritten as
\begin{linenomath}
\begin{equation}
  	n_l^c(t+1) = n_l^c(t) + \frac{1}{L_l} \left( \sum_{i=1}^M f_{il}^c(t) - \sum_{j=1}^N f_{lj}^c(t) \right), \label{eq:ctm_multiclass}
\end{equation}
\end{linenomath}
where $c \in \{1,\dots,C\}$ indexes vehicle classes (often called ``commodities'' in the traffic literature).

Extending the density update equation to multiple classes means that the link and node models must also be extended to produce per-class flows $f_{ij}^c$.
In this paper, we will not specify a particular link model, but assume use of one that produces per-class demands $S_i^c$ and overall supplies $R_j$ (the node model, in computing the $f_{ij}^c$, is responsible for splitting the available supply $R_j$ among the different demanding vehicle classes).
Examples of this type of link model include those considered in \citet{wong_multi-class_2002}, \citet{daganzo_behavioral_2002}, and \citet{van_lint_fastlane_2008} (examples of multi-class link models of second- or higher-order include those of \citet{hoogendoorn_continuum_2000}. These types of models have higher-order analogs of supply and demand).

\subsection{Toplogical expression of managed lane-freeway networks}\label{subsec_topological_expression}
In both~\eqref{eq:ctmbasic} and~\eqref{eq:ctm_multiclass}, we describe a link in terms of its total density $n_l$ and its breakdown into per-commodity portions, $n_l^c$.
By discretizing the road into these one-dimensional links, we lose information about differences between vehicle proportions across lanes, as well as inter-lane and lane-changing behavior.
This becomes a problem if such unmodeled behavior is of interest.
In our setting, this means that modeling a freeway with a managed lane should not be done with a single link following~\eqref{eq:ctmbasic} or~\eqref{eq:ctm_multiclass}, as it would be impossible to study the managed lane-freeway network behavior of interest.

Modeling differences in vehicle density across lanes is natural in microscopic and mesoscopic (see, for example, \citet{treiber_derivation_1999, hoogendoorn_gas-kinetic_1999, ngoduy_macroscopic_2006}, and others) models, but macroscopic models, in their simplicity, have less readily-accessible avenues for including these differences.
One straightforward method is to model each lane as a separate link, as in, e.g. \citet{bliemer_dynamic_2007} or \citet{shiomi_multilane_2015}.
However, this method has a few drawbacks.
First, it requires the addition of some lane-assignment method to prescribe the proportions of each vehicle class $c$ for each lane (such as a logit model as used in \citet{farhi_logit_2013} and \citet{shiomi_multilane_2015}), which requires not-always-accessible data for calibration.
Second, drastically increasing the number of links in a macroscopic model will necessarily increase the size of the state space and model complexity, which is, in a sense, incompatible with the overall goal of selecting a macroscopic model over a micro- or mesoscopic model: some of the ``macro'' in the macroscopic model is lost. 

Instead, in this paper we choose to model the GP lanes (or ``GP lane group'') as one link and the parallel managed lanes (or ``managed lane group'') as another link.
Applied to an entire length of road, this creates a network topology of two ``parallel chains'' of links - one GP and one managed.
The two chains will share nodes, but cross-flows between the chains are permitted only in locations where there is physical access (i.e., no physical barriers) and policy access (i.e., no double solid lines under U.S. traffic markings).
Where cross-flows are possible, we do not use a logit model, but instead a driver behavior model first introduced in \citet{wright_node_2017}.
This two-chain model is similar to the one described in \citet{liu_modeling_2012}, though in this reference, GP-managed lane crossflows were not considered.

\subsection{Modeling emergent phenomena particular to managed lane-freeway networks}\label{subsec_observedphenoms}
Our techniques for modeling the friction, smoothing, and inertia effects are rooted in a full-picture view of the macroscopic modeling framework.
As we will see, we propose isolating each of these ``effects'' to one component of the traditional macroscopic modeling framework: the friction effect model is based on a principled feedback mechanism of the fundamental diagram, the inertia effect model is included in modeling of driver choices for which of the two lane groups they will take, and the smoothing effect model emerges from a particular recently-developed node model construction.

\newcommand{\obeta}{\overline{\beta}}
\newcommand{\tbeta}{\tilde{\beta}}

\section{Full- and Gated-Access Managed Lane-Freeway Network Topologies}\label{sec_hov_model}
We will consider two types of managed lane-freeway network configurations:
\emph{full access} and \emph{separated with gated access}.
In a a full-access configuration, the managed lane(s) are not physically separated from the GP lane(s), and eligible vehicles may switch between the two lane groups at any location.
Often, full-access managed lane(s) are special-use only during certain periods of the day, and at other times they serve as GP lane(s) (e.g., HOV lanes are often accessible to non-HOVs outside of rush hour).
On the other hand, in a gated-access configuration, traffic may switch between the managed lane(s) and GP lane(s) only at certain locations, called \emph{gates}; at non-gate locations, the two lane groups are separated by road markings (i.e., a double solid line in the U.S.) or a physical barrier.
Usually, gated-access managed lanes are special-use at all times.
The implemented managed lane access scheme depends on jurisdiction.
For example, full-access lanes are common in Northern California,
and separated lanes are common in Southern California.

The differences in physical geometry and access points between the two access types requires two different types of topology in constructing a network for a macroscopic model.

\subsection{Note on link and node models used in this section}\label{subsec_link_node_model_note}
As discussed in the previous Section, we attempt to be agnostic with regards to the particular link model (e.g., first-order fundamental diagram) used in our implementations.
However, in our model of the friction effect in Section~\ref{subsec_friction}, we parameterize friction being in effect on a particular link $l$ for a particular vehicle class $c$ at time $t$ by adjusting that link's demand $S_l(t)$.
For that discussion only, we specify a particular link model.
This link model is reviewed in Appendix~\ref{app_linkmodel}.

The node model used here and for the remainder of this paper, when a particular form is necessary, is the one discussed in \citet{wright_dynamic_2016,wright_node_2017}.
We use this node model because it handles multi-commodity traffic, optimizes the utilization of downstream supply, makes use of input link priorities and has a relaxation of the ``conservation of turning ratios'' or ``first-in-first-out'' (FIFO) constraint of most node models.
This last feature allows us to describe a set of GP lanes just upstream of an offramp with one link, and handle a condition of a congested offramp by having the congestion spill back onto only the offramp-serving lanes of the GP link (as opposed to the entirety of the GP link).
See \citet[Section 3]{wright_node_2017} and \citet{wright_dynamic_2016} for more discussion.

\subsection{Full-access managed lanes}
\label{subsec_full_access}
\begin{figure}[tb]
\centering
\includegraphics[width=\textwidth]{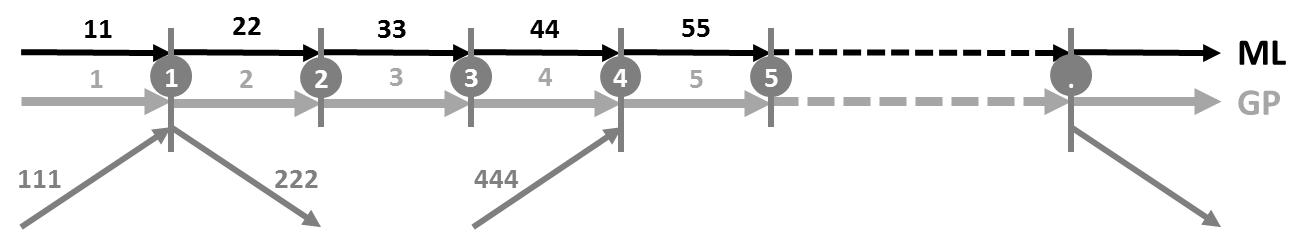}
\caption{Freeway with full-access managed lane. ML = Managed Lane.}
\label{fig-full-access-hov}
\end{figure}

A full-access managed lane configuration is presented in
Figure~\ref{fig-full-access-hov}:
GP and managed links (recall, as discussed above, all GP lanes and all managed lanes are collapsed into one link each) are parallel with the same geometry
and share the same beginning and ending node pairs;
traffic flow exchange between GP and managed lanes can happen at every node.
Note that in Figure~\ref{fig-full-access-hov}, we use a slightly irregular numbering scheme so that it is clear whether a link is a GP link, managed lane link, or ramp link.
Parallel links in the graph have numbers made up of the digit of their terminating node, with GP links having one digit (i.e., link 1), managed lane links having two (i.e., link 11), and ramp links having three (i.e., link 111).
Note also that we use a U.S.-style, driving-on-the-right convention here, with the managed lane(s) on the left of the GP lanes and the ramps on the far right.

Links that are too long for modeling purposes (i.e., that create too low-resolution a model) may be broken up into smaller ones by creating
more nodes, such as nodes 2 and 3 in Figure~\ref{fig-full-access-hov}.
Fundamental diagrams for parallel GP and managed lane links may be different~\citep{liu_analysis_2011}.

We introduce two vehicle classes ($C=2$): 
$c=1$ corresponds to the GP-only traffic and
$c=2$ corresponds to the special traffic.
When the managed lane(s) is (are) active, $c=1$-traffic is confined to the GP link,
whereas $c=2$-traffic can use both the GP and managed lane links.
We denote the portion of vehicles of class $c$ in link $i$ that will attempt to enter link $j$ as $\beta_{i,j}^c$.
This quantity is called the split ratio.

\subsubsection{Split ratios for full-access managed lanes}\label{subsubsec_fullaccess_splits}
We make an assumption that both vehicle classes take offramps at the same rate.
For example, for node 1 in Figure~\ref{fig-full-access-hov}, we might say that $\beta_{i,222}^1 = \beta_{i,222}^2 \triangleq \beta_{i,222}$.
Strictly speaking, it is not necessary to assume that the $\beta_{i,222}^c$ are equal for all $c$.
However, in practice the offramp split ratios are typically estimated from flow count data taken from detectors on the offramp and freeway.
Generally speaking, these detectors cannot identify vehicle type, so the only quantity estimable is a flow-weighted average of the quantities $\beta_{i,222}^c$.
To estimate the class-specific split ratios, one needs some extra knowledge of the tendency of each class to take each offramp (for example, that GP-only vehicles are half as likely as special traffic to take a certain offramp).
Assuming that each class exits the freeway at the same rate is a simple and reasonable-seeming assumption.

This same problem of unidentifiability from typical data appears in several other split ratios.
First, it may not be possible to tell how many vehicles taking an offramp link come from the upstream GP, managed lane, or (if present) onramp link.
In this case, some assumptions must then be made.
For example, three different assumptions that may be reasonable are (1) that vehicles in each link $i$ take the offramp at the same rate; or (2) that no vehicles from the managed lane(s) are able to cross the GP lanes to take the offramp at this node, and that no vehicles entering via the onramp, if one is present, exit via the offramp at the same node; or
(3) that vehicles in GP and managed links take the offramp at the same rate, while no vehicles coming from
the onramp are directed to the offramp.
Looking back at node 1 in Figure~\ref{fig-full-access-hov} again, assumption (1) would say the $\beta_{i,222}$ are equal for all $i$; assumption (2) would say $\beta_{11,222}=\beta_{111,222} = 0$;
and assumption (3) would say $\beta_{1,222}=\beta_{11,222}$ and $\beta_{111,222}=0$.
The best assumption for each node will depend on the road geometry for that particular part of the road (how near the offramp is to any onramps, how many GP lanes a vehicle in a managed lane would have to cross, etc.).

Second, the crossflows between the GP and managed lane links are not observable.
Even if there exist detectors immediately upstream and downstream of the node where traffic can switch between
GP and managed lanes, it is impossible to uniquely identify the crossflows.
In a simulation, these crossflows must be governed by some driver choice model.

Putting together these assumptions and the special-traffic-only policy for the managed lane, we can summarize most of the necessary split ratios needed for computing flows in a node model.
For example, for node 1 in Figure~\ref{fig-full-access-hov},
\begin{linenomath}
\begin{align*}
\beta_{i,j}^1 &= \left\{ 
\begin{array}{ l l l l }
  \multicolumn{1}{r}{} & \multicolumn{1}{c}{\bm{j=2}} & 
    \multicolumn{1}{c}{\bm{ j = 22}} & \multicolumn{1}{c}{\bm{ j = 222}} \\
  \cline{2-4}
  \bm{i = 1} & 1 - \beta_{1,222} & 0 & \beta_{1,222} \\
  \bm{i = 11} & \text{n/a} & \text{n/a} & \text{n/a} \\
  \bm{i = 111} & 1 - \beta_{111,222} & 0 & \beta_{111,222} \\
\end{array} \right. \\
\beta_{i,j}^2 &= \left\{ 
\begin{array}{ l l l l }
  \multicolumn{1}{r}{} & \multicolumn{1}{c}{\bm{j=2}} & 
    \multicolumn{1}{c}{\bm{ j = 22}} & \multicolumn{1}{c}{\bm{ j = 222}} \\
  \cline{2-4}
  \bm{i = 1} & - & - & \beta_{1,222} \\
  \bm{i = 11} & - & - & \beta_{11,222} \\
  \bm{i = 111} & - & - & \beta_{111,222} \\
\end{array} \right.
\end{align*}
\end{linenomath}
where ``n/a'' means that the split ratios $\beta_{11,j}^1$ are not applicable, as there should be no vehicles of class $c=1$ in the managed lane.
The split ratios marked with a dash are those above-mentioned flows that are unobservable and come from some driver choice model.
Of course, whatever method is chosen to compute these unknown split ratios, we must have $\beta_{i,2}^2 + \beta_{i,22}^2 = 1 - \beta_{i,222}^2$.

Similarly, for node 2, which does not have an onramp or an offramp,
\begin{linenomath}
\begin{align*}\beta_{i,j}^1 &= \left\{ 
\begin{array}{ l l l }
  \multicolumn{1}{r}{} & \multicolumn{1}{c}{\bm{j=3}} & 
    \multicolumn{1}{c}{\bm{ j = 33}} \\
  \cline{2-3}
  \bm{i = 2} & 1 & 0 \\
  \bm{i = 22} & \text{n/a} & \text{n/a} \\
\end{array} \right. \\
\beta_{i,j}^2 &= \left\{ 
\begin{array}{ l l  l}
  \multicolumn{1}{r}{} & \multicolumn{1}{c}{\bm{j=3}} & 
    \multicolumn{1}{c}{\bm{ j = 33}} \\
  \cline{2-3}
  \bm{i = 2} & - & - \\
  \bm{i = 22} & - & - \\
\end{array} \right.
\end{align*}
\end{linenomath}
with ``n/a'' and the dash meaning the same as above.

As previously mentioned, full-access managed lanes often have certain time periods during which nonspecial ($c=1$) traffic is allowed into the managed lane.
We can model this change in policy simply by changing the split ratios at the nodes.
For node 1, for example, the nonrestrictive policy is encoded as
\begin{linenomath}
\begin{align*}
\beta_{i,j}^c &= \left\{ 
\begin{array}{ l l l l }
  \multicolumn{1}{r}{} & \multicolumn{1}{c}{\bm{j=2}} & 
    \multicolumn{1}{c}{\bm{ j = 22}} & \multicolumn{1}{c}{\bm{ j = 222}} \\
  \cline{2-4}
  \bm{i = 1} & - & - & \beta_{1,222} \\
  \bm{i = 11} & - & - & \beta_{11,222} \\
  \bm{i = 111} & - & - & \beta_{111,222} \\
\end{array} \right. \qquad \text{for } c = \{1,2\},
\end{align*}
and for node 2 as
\begin{align*}
\beta_{i,j}^c &= \left\{ 
\begin{array}{ l l l }
  \multicolumn{1}{r}{} & \multicolumn{1}{c}{\bm{j=3}} & 
    \multicolumn{1}{c}{\bm{ j = 33}} \\
  \cline{2-3}
  \bm{i = 2} & - & - \\
  \bm{i = 22} & - & - \\
\end{array} \right. \qquad \text{for } c = \{1,2\}.
\end{align*}
\end{linenomath}
In other words, the managed lane link is treated as additional GP lane(s), and the split ratios governing the crossflows between the two links should be found from the driver choice model for both vehicle classes.

\subsubsection{Node model for full-access managed lane-freeway networks}\label{subsubsec_fullaccess_node}
As mentioned in Section \ref{subsec_link_node_model_note}, we make use of the node model discussed in~\citet{wright_dynamic_2016,wright_node_2017} to describe a freeway network with managed lanes.
This node model differentiates itself from others in that it deals with multi-commodity traffic flow,
optimally utilizes the available supply,
makes use of input link priorities, and
has a relaxation of the common FIFO constraint.
By default, link priorities can be taken proportional to link capacities.
To explain relaxed FIFO, say that some link $i$ has vehicles that wish to enter both links $j$ and $j'$.
If link $j'$ is jammed and cannot accept any more vehicles, a strict FIFO constraint would say that the vehicles in $i$ that wish to enter $j'$ will queue at $i$'s exit, and block the vehicles that wish to enter $j$.
In a multi-lane road, however, only certain lanes may queue, and traffic to $j$ may still pass through other lanes.
The relaxation is encoded in so-called ``mutual restriction intervals'' $\etab_{j'j}^i \subseteq [0,1]$.
This interval partly describes the overlapping regions of link $i$'s exit that serve both links $j$ and $j'$.
For $\etab_{j'j}^i=[y,z]$, a $z-y$ portion of $i$'s lanes that serve $j$ also serve $j'$, and will be blocked by the cars queueing to enter $j'$ when $j'$ is congested.
For example, if $j$ is served by three lanes of $i$, and of those three lanes, the leftmost also serves $j'$, we would have $\etab_{j'j}^i = [0, \sfrac{1}{3}]$.

As an example, we consider again node 1 in Figure \ref{fig-full-access-hov}.
Say that the GP links (1 and 2) have four lanes, that the managed lane links (11 and 22) have two lanes, and that the onramp merges into and the offramp diverges from the rightmost GP lane.
Further, we say that when the managed lane link is congested, vehicles in the GP lanes that wish to enter the managed lanes will queue only in the leftmost GP lane.
On the other hand, when the GP link is congested, vehicles in the managed lanes that wish to enter the GP lanes will queue only in the rightmost managed lane.
Finally, we suppose that jammed offramp (222) will cause vehicles to queue only in the rightmost GP lane.
Taking together all of these statements, our mutual restriction intervals for this example are:
\begin{linenomath}
\begin{align*}
\etab_{j'j}^1 &= \left\{ 
\begin{array}{ l c c c }
  \multicolumn{1}{r}{} & \multicolumn{1}{c}{\bm{j=2}} & 
    \multicolumn{1}{c}{\bm{ j = 22}} & \multicolumn{1}{c}{\bm{ j = 222}} \\
  \cline{2-4}
  \bm{j' = 2} & [0,1] & [0,1] & [0,1] \\
  \bm{j' = 22} & \left[0, \sfrac{1}{4} \right] & [0,1] & \emptyset \\
  \bm{j' = 222} & \left[ \sfrac{3}{4}, 1 \right] & \emptyset & [0,1] \\
\end{array} \right. \\
\etab_{j'j}^{11} &= \left\{ 
\begin{array}{ l c c c }
  \multicolumn{1}{r}{} & \multicolumn{1}{c}{\bm{j=2}} & 
    \multicolumn{1}{c}{\bm{ j = 22}} & \multicolumn{1}{c}{\bm{ j = 222}} \\
  \cline{2-4}
  \bm{j' = 2} & [0,1] & [0,\sfrac{1}{2}] & \emptyset \\
  \bm{j' = 22} & \left[0, 1\right] & [0,1] & \emptyset \\
  \bm{j' = 222} & \emptyset & \emptyset & [0,1] \\
\end{array} \right. \\
\etab_{j'j}^{111} &= \left\{ 
\begin{array}{ l c c c }
  \multicolumn{1}{r}{} & \multicolumn{1}{c}{\bm{j=2}} & 
    \multicolumn{1}{c}{\bm{ j = 22}} & \multicolumn{1}{c}{\bm{ j = 222}} \\
  \cline{2-4}
  \bm{j' = 2} & [0,1] & [0,1] & [0,1] \\
  \bm{j' = 22} & \emptyset & [0,1] & \emptyset \\
  \bm{j' = 222} & [0,1] & [0,1] & [0,1] \\
\end{array} \right. .
\end{align*}
\end{linenomath}
To read the above tables, recall that as written, $j'$ is the congested, restricting link, and $j$ is the restricted link.
These chosen restriction intervals allow for expected behavior in this network, such as a congested GP link causing possible queueing in the right managed lane (if some drivers are trying to enter the GP link), but no spillback into the left managed lane.\footnote{In this example, we assume that the managed lane
has two sublanes --- left and right.}

For a detailed discussion on how mutual restriction intervals are included in the node model's flow calculations and solution algorithms, see~\citet{wright_dynamic_2016,wright_node_2017}.

\subsection{Separated managed lanes with gated access}\label{subsec_separated}
\begin{figure}[htb]
\centering
\includegraphics[width=\textwidth]{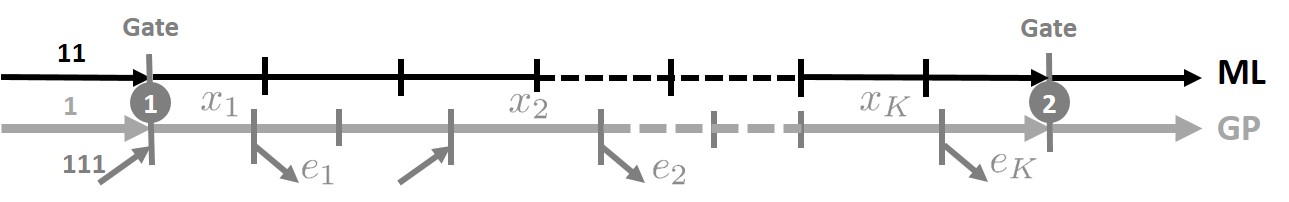}
\caption{Freeway with separated managed lane and gates. ML = Managed Lane.}
\label{fig-separated-hov}
\end{figure}

A separated, gated-access managed lane configuration
is presented in Figure~\ref{fig-separated-hov}.
Unlike the full-access configuration, the GP and managed lane link chains do not necessarily meet at every node.
Instead, they need only meet at a few locations, where vehicles can move into and out of the managed lane(s).
Note that, unlike the full-access configuration, there is no need for GP and managed lane links to be aligned.

As labeled in Figure~\ref{fig-separated-hov}, the nodes where the two link chains meet are called \emph{gates} (as an aside, one way to describe the full-access managed lane configuration would be that every node is a gate).
Similar to our construction of excluding GP-only traffic from the managed lane in the full-access case, we can disable flow exchange at a given gate by fixing split ratios
so that they keep traffic in their lanes.
For example, to disable the gate
(the flow exchange between the two lanes) at node~2
in Figure~\ref{fig-full-access-hov}, we set
$\beta_{2,3}^c=1$ and $\beta_{22,33}^c=1$
(this means that $\beta_{2,33}^c=0$ and $\beta_{22,3}^c=0$), $c=1,2$.
Thus, the full-access managed lane can be easily converted
into the separated managed lane by setting non-exchanging split ratios everywhere
but designated gate-nodes.

In practice, a gate is stretch of freeway that may be a few hundreds of meters long \citep{cassidy_problem_2015},
and, potentially, we can designate two or three sequential nodes as gates.
In this paper, however, we model a gate as a single node.

For the gated-access configuration, we suggest setting mutual restriction
coefficients in the same manner as full-access managed lanes, in Section \ref{subsubsec_fullaccess_node}.

Compared to the full-access managed lane configuration, the gated-access configuration
has a smaller friction effect~\citep{jang_traffic_2012}: drivers in the
separated managed lane feel somewhat protected
by the buffer, whether it is virtual (double solid line) or real (concrete),
from vehicles changing abruptly from the slow moving GP lane and, therefore,
do not drop speed as dramatically.
The degree to which the friction effect
is mitigated is disputed (e.g., see Footnote 3 in~\citet{cassidy_problem_2015}),
but overall the bottlenecks created by the gates are
much greater instigators of congestion~\citep{cassidy_problem_2015}.
Inclusion of the friction effect in modeling
separated managed lane configurations is thus not as essential as in modeling
the full-access case.

By similar logic, the smoothing effect should be expected to be less prominent.
If a physical barrier prevents lane changing into the managed lane, then there will be no lane changing into the managed lane - the flow is already as ``smoothed'' as possible.
It will become apparent in section \ref{sec:smoothing}, when we discuss our smoothing effect model, that in our model the degree of traffic smoothing is indeed maximized when the inter-link split ratios are zero (as they are in the non-gate nodes).

\subsubsection{Modeling a flow of vehicles from the managed lanes to the offramps}\label{subsubsec_pipeclasses}
Recall from Section~\ref{subsubsec_fullaccess_splits} that, in the full-access managed lane model, we can model vehicles moving from the managed lane link to offramps in a straightforward manner, by setting corresponding split ratios (for example,
$\beta_{11,222}^c$, $c=1,2$, for node~1 in configuration from
Figure~\ref{fig-full-access-hov}).
For the gated-access configuration, however, modeling traffic as moving from the managed lanes to offramps is more complicated:
generally, gates do not coincide with offramp locations.
In fact, there are typically between two and five offramps between two gate locations.
These offramps cannot be accessed directly from the managed lane.
To model a vehicle flow that originates at the managed lane link, goes through a gate to the GP lane link, and then takes the correct offramp (and doing so at some rate that matches external origin-destination data), requires a more involved modeling and bookkeeping construction.

To resolve this challenge, our gated-access model introduces new vehicle classes
in addition to the $c=1$ (GP-only) and $c=2$ (special) traffic used in the full-access model of Section~\ref{subsec_full_access}.
These additional classes will be used to distinguish subsets of the special traffic population
by its destination offramp.
If $K$ is the largest number of offramps
between two adjacent gates, then altogether we have $C=K+2$ vehicle classes:
$c=1, 2, e_1, \dots, e_K$, where $e_k$ indicates the class of vehicles that
will exit through the $k$-th offramp after leaving the managed lane through the gate.
By definition, traffic of type $c=e_k$ \emph{may} exist in the GP lane segment
between gate~1 and offramp $e_k$, but there is \emph{no traffic of this
type} either in the GP link segment between offramp $e_k$ and gate~2 or
in the managed lane link.
This movement pattern is ensured by setting \emph{constant} split ratios:
\begin{linenomath}
\begin{equation}
\begin{array}{ll}
\beta_{i \; x_1}^{e_k} = 1, \;\;\; i=1,11,111, &
\mbox{ direct all $e_k$-type traffic to the GP link at gate~1};\\
\beta_{x_k e_k}^{e_k} = 1, &
\mbox{ direct all $e_k$-type traffic to offramp $e_k$};\\
\beta_{x_{k'}e_{k'}}^{e_k} = 0, \;\;\; k'\neq k, &
\mbox{ do not send any $e_k$-type traffic to other offramps},
\end{array}
\label{eq_destination_sr}
\end{equation}
\end{linenomath}
where $k=1,\dots,K$, and $x_k$ denotes the input GP link for the node that
has the output link $e_k$ (see Figure~\ref{fig-separated-hov}).

\newcommand{\nt}{\tilde{n}}

Vehicles of class $c=e_k$ do not enter the network via onramps or the upstream boundary, but instead are converted from special $c=2$ traffic as it leaves the managed lane(s) through the gate.
We perform this conversion as part of the link model computation, such that the total demand for the switching link, $\sum_{c=1}^C S_{l}^c$ remains the same before and after the switch.
The exact link on which this switching takes place is the managed lane link immediately upstream of the gate (e.g., link 11 in Figure~\ref{fig-separated-hov}).

We say that the amount of traffic that should change from vehicle class $c=2$ to vehicle class $e_k$ at time $t$ is (using link 11 as an example) $n_{11}^2(t) \beta_{x_k e_k}^2(t) v_{11}(t)$, where $v_{11}(t)$ is in units of vehicles per simulation timestep (this factor is included so that the switching done is proportional to link 11's outflow, rather than its density), and $\beta_{x_k e_k}^2(t)$ is the split ratio from $x_k$, the GP link immediately upstream of exit $e_k$, and the exit $e_k$ at time $t$.
This statement is based on the assumption that the vehicles in the managed lane link will exit the freeway through exit $e_k$ at the same rate as special ($c=2$) vehicles that happened to stay in the GP lanes.
That is, if a $\beta_{x_k e_k}^2$ portion of $c=2$ vehicles intend to leave the GP lanes through exit $e_k$, then a $\beta_{x_k e_k}^2$ portion of the $c=2$ vehicles in the managed lane(s) will leave the managed lane link at the closest upstream gate and leave the network at exit $e_k$ when they reach it.
Note that if there are $K' < K$ offramps between two particular gates, then no vehicles should switch to type $c=e_{k}, k \in \{K'+1,\dots,K\}$ at the upstream gate, as they would have no ramp to exit through.

We now explain how $e_k$-type traffic appears in the system.
The original demand $d_l^c(\cdot)$ is specified at origin links $l$ for
commodities $c=1,2$, and $d_l^{e_k}(\cdot)\equiv 0$, $k=1,\dots,K$.
Destination specific traffic appears in the managed lane links that end at gate-nodes
by assigning destinations to portions of the type-1 (GP-only) and type-2 (managed lane-eligible)
traffic in those links.
We propose using offramp split ratios $\beta_{x_k,e_k}^c$, $c=1,2$, $k=1,\dots,K$,
to determine portions of managed lane traffic to be assigned particular 
destinations.
The destination assignment algorithm at a given time $t$,
for a given HOV link ending with a gate-node, is described next.
Without loss of generality, we will refer to
Figure~\ref{fig-separated-hov} and managed lane link~11 ending at the gate-node~1
in this description.
Using Figure~\ref{fig-separated-hov} as a reference, we can now formally
describe the procedure for destination assignment to traffic in the
managed lane link.
\begin{enumerate}
\item Given are vehicle counts per commodity $n_{11}^{c}$,
$c=1,2,e_1,\dots,e_K$; free flow speed $v_{11}$; and offramp
split ratios $\beta_{x_k,e_k}^1$ and $\beta_{x_k,e_k}^2$,
$k=1,\dots,K$.\footnote{If a given GP segment connecting two adjacent gates
has $K'$ offramps, where $K'<K$, then assume
$\beta_{x_k,e_k}^1=\beta_{x_k,e_k}^2=0$ for $k\in(K', K]$.}

\item Initialize:
\begin{eqnarray*}
\nt_{11}^c(0) & := & n_{11}^c, \;\;\; c=1,2,e_1,\dots e_K;\\
k & := & 1.
\end{eqnarray*}

\item Assign $e_k$-type traffic:
\begin{eqnarray}
\nt_{11}^{e_k}(k) & = & \nt_{11}^{e_k}(k-1) + 
\beta_{x_k,e_k}^1 v_{11} \nt_{11}^1(k-1) +
\beta_{x_k,e_k}^2 v_{11} \nt_{11}^2(k-1); \label{eq_commodity_ek}\\
\nt_{11}^{1}(k) & = & \nt_{11}^{1}(k-1) -
\beta_{x_k,e_k}^1 v_{11} \nt_{11}^1(k-1); \label{eq_commodity_1}\\
\nt_{11}^{2}(k) & = & \nt_{11}^{2}(k-1) -
\beta_{x_k,e_k}^2 v_{11} \nt_{11}^2(k-1). \label{eq_commodity_2}
\end{eqnarray}

\item If $k<K$, then set $k:=k+1$ and return to step~3. 

\item Update the state:
\[
n_{11}^c = \nt_{11}^c(K), \;\;\; c=1,2,e_1,\dots,e_K.
\]
\end{enumerate}

After the switches to $c=e_k$ class traffic have been done, there may be unresolved split ratios for both classes $c=1$ and $c=2$ at the gates (similar to the dashed split ratios in the tables in Section \ref{subsubsec_fullaccess_splits}).
These split ratios should be filled in with the same tools as those in Section \ref{subsubsec_fullaccess_splits}: some sort of driver lane choice behavior.

\section{Modeling Emergent Phenomena Particular to Managed Lane-Freeway Networks}\label{sec_phenomena}
As previously discussed, managed lane-freeway networks exhibit macroscopic phenomena that do not arise in situations where the managed lanes may act as traditional GP lanes
\footnote{One of these effects, the smoothing effect, was even observed in a previous study \citep{cassidy_smoothing_2010} to be present in a managed lane-freeway network, but \emph{not present on the same section of road} in another period of the day when the managed lane policy was not enforced (in that particular case, carpool/HOV lane enforcement during the peak hours)}.
These unique behaviors are caused by the interactions between the qualitatively different the flows in the GP lanes and the neighboring managed lanes.
In this section, we describe simple physical models for three such behaviors: the friction effect, the smoothing effect, and the inertia effect.

\subsection{Friction effect}\label{subsec_friction}
The \emph{friction effect} is an empirically-observed phenomenon in situations where managed lanes are relatively uncongested, but the managed-lane traffic will still slow down when the adjacent GP lanes congest and slow down (see \citet{daganzo_effects_2008,liu_analysis_2011,thomson_operational_2012, fitzpatrick_operating_2017}, etc.).
It has been hypothesized \citep{jang_dual_2012} that this phenomenon arises from the managed-lane drivers' fear that slower-moving vehicles will suddenly and dangerously change into the managed lane ahead of them.

We suggest modeling the friction effect
based on a \emph{feedback mechanism} that uses the difference
of speeds in the parallel GP and managed lane links to scale down the flow (and therefore the speed)
out of the managed lane link if necessary.

To explain the concept, we again refer to Figure~\ref{fig-full-access-hov}
and consider parallel links 1 (GP) and 11 (managed lane).
Recall that, under a first-order model \eqref{eq:ctm_multiclass}, the speed of traffic in link $l$ at time $t$ is
\begin{linenomath}
\begin{equation}
  v_l(t) = \begin{cases}
    \frac{ \sum_{c=1}^C \sum_{j=1}^N f_{lj}^c(t)}{ \sum_{c=1} n_l^c(t)} & \text{if } \sum_{c=1}^c n_l^c(t) > 0, \\
    v_l^f(t) & \text{otherwise},
  \end{cases}
  \label{eq_speed_formula}
\end{equation}
\end{linenomath}
where $v_l^f(t)$ is the theoretical free flow speed of link $l$ at time $t$.

We say that the friction effect is present in managed lane link 11
(following the notation of Figure~\ref{fig-full-access-hov}) at time $t$ if
\begin{linenomath}
\begin{equation}
v_1(t-1) < \min\left\{v_1^f,\; v_{11}(t-1)\right\},
\label{eq_friction_condition}
\end{equation}
\end{linenomath}
which means that
(1) the GP link is in congestion (its speed is below its current free flow speed), and
(2) the speed in the GP link is less than the speed in the managed lane link.
We denote this speed differential as:
\begin{linenomath}
\begin{equation}
\Delta_{11}(t) = v_{11}^f - v_1(t-1).
\label{eq_delta}
\end{equation}
\end{linenomath}

It has been observed \citep{jang_traffic_2012} that the magnitude of the friction effect --- the degree to which managed-lane drivers slow down towards the GP lane's traffic speed --- depends on the physical configuration of the road.
For example, less of a friction effect will be present on managed lanes that are separated from the GP lanes by a buffer zone than those that are contiguous with the GP lanes \citep{thomson_operational_2012, jang_traffic_2012,fitzpatrick_operating_2017}, and the presence of a concrete barrier would practically eliminate the friction effect.
Other factors that may affect this magnitude include, for example, whether there is more than one managed lane, or whether there is a shoulder lane to the left of the managed lane that drivers could swerve into if necessary.

To encode this variability in the magnitude of the friction effect in managed lane link 11, we introduce 
$\sigma_{11}\in [0, 1]$ the \emph{friction coefficient} of this link.
The friction coefficient reflects the strength of the friction.
Its value depends on the particular managed configuration and is chosen by the modeler.
A value of $\sigma_{11}=0$ means
there is no friction (which may be appropriate if, perhaps, the managed lane(s) are separated from the GP lanes by a concrete barrier), and $\sigma_{11}=1$ means that the managed lane link speed tracks the GP link speed exactly.

When the friction effect is active (i.e., when~\eqref{eq_friction_condition} is true), we adjust the fundamental diagram of the managed lane link by scaling down its theoretical free flow speed $v_l^f(t)$, and propagate that change through the rest of the fundamental diagram parameters.
The exact mathematical changes will of course be different for every different form of fundamental diagram.
For the particular fundamental diagram discussed in Appendix~\ref{app_linkmodel}, this means adjusting the free flow speed and capacity as follows:
\begin{linenomath}
\begin{align}
\hat{v}_{11}(t) & = v_{11}^f(t) - \sigma_{11}\Delta_{11}(t);
\label{eq_friction_free_flow_speed} \\
\hat{F}_{11}(t) & = \hat{v}_{11}(t)n_{11}^+, \label{eq_friction_capacity}
\end{align}
\end{linenomath}
where $n_{11}^+$ is the high critical density
(see Appendix~\ref{app_linkmodel} for its definition),
and using these adjusted values in the calculation of the sending
function~\eqref{eq_lnctm_send_function},
\begin{linenomath}
\begin{equation}
S_{11}^c(t) =
\hat{v}_{11}(t)n_{11}^c(t)\min\left\{1,
\frac{\hat{F}_{11}(t)}{\hat{v}_{11}(t)\sum_{c=1}^C n_{11}^c(t)}\right\}.
\label{eq_friction_send_function}
\end{equation}
\end{linenomath}
For the fundamental diagram of Appendix \ref{app_linkmodel}, we must also check whether 
\begin{linenomath}
\begin{equation}
\sum_{c=1}^C n_{11}^c(t) <
\frac{\hat{F}_{11}(t)}{v_{11}^f - \Delta_{11}(t)}
= \frac{\hat{v}_{11}(t)n_{11}^+(t)}{v_{11}^f - \Delta_{11}(t)}.
\label{eq_friction_threshold}
\end{equation}
\end{linenomath}
If not, then applying friction will lead to the managed lane link speed falling below the GP link speed, and the unadjusted sending function should be used (the possibility of this happening is due to the use of two critical densities to create Appendix \ref{app_linkmodel}'s fundamental diagram's ``backwards lambda'' shape).

Again, the exact form of \eqref{eq_friction_send_function}, the sending function with friction, will depend on the link's original fundamental diagram model.

\subsubsection{Comparison with related work}
We are not the first to proposed modeling the friction effect by adjusting the fundamental diagram of the managed lane as a function of the state of the GP lane(s).
In particular, \citet{liu_modeling_2012} consider a stretch of an HOV-equipped freeway where the two classes of lanes are separated by a buffer, and propose to fit two different sets of flow parameters for the HOV lane based on whether or not the GP lanes are congested.

Our method differs in formulation and potential uses.
First, we propose a principled modification to the fundamental diagram that is based on observations of how the friction effect actually affects the managed lane's fundamental diagram, via the idea of lowering the theoretical managed lane free-flow speed.
Examining \eqref{eq_friction_free_flow_speed} again, this formulation has an effect on the fundamental diagram's free-flow speed that is linear in the speed differential.
This formulation was chosen based on prior studies: evidence of a friction effect reducing managed lane speed in a form that is linear in the speed differential was reported in, e.g., \citet{jang_traffic_2012}.
This is in contrast to earlier models that switch between two distinct and discrete sets of fundamental diagram parameters based on a GP congestion threshold (i.e., a piecewise rather than linear relationship).

Our theoretically-based fundamental diagram adjustment is also motivated by the common use for macroscopic simulation as a forecasting tool.
Traffic simulation tools are often used to study potential effects of proposed, but not yet developed, infrastructure construction~\citep{kurzhanskiy_traffic_2015}.
Moreover, the particular infrastructure investment of managed lane construction introduces the potential for additional policy choices (e.g., whether to restrict an HOV lane to vehicles with at least 2 occupants or those with at least 3 occupants, the tolling prices in a tolled express lane, etc.), each with their own feedback effects of the flow on the road, and underlying demand management.
These complexities have motivated the development of new simulation tools specifically for new managed lane types \citep{guohui_simulation-based_2009}.
In the business and policy worlds, these complex investments with many feedback effects are evaluated through Monte Carlo simulation of principled models \citep{kurzhanskiy_traffic_2015}.
Our friction effect model is motivated in part by this type of use case, where a theoretically-justified model is to be used in situations where data is unavailable due to the construction having not occurred yet.
On the other hand, of course, if a traffic engineer is studying an already-existing managed lane-freeway network, where data measuring the friction effect can be collected, it makes sense to fit a friction effect model based on those data.

\subsection{Inertia effect}\label{subsec_inertia}
At the end of Section \ref{subsubsec_pipeclasses}, we mentioned that some split ratios will likely be undefined after switching $c=2$ class vehicles to the $c=e_k$ classes.
In particular, $c=2$ class vehicles in the upstream GP link (e.g., link 1 in Figure \ref{fig-separated-hov}) and remaining $c=2$ vehicles in the upstream managed lane link (e.g. link 11 in Figure \ref{fig-separated-hov}) will need to decide whether to pass through the gate or remain in their current lane.
Sample tools for modeling driver lane choices such as these include the class of ``logit'' logistic regression models \citep{mcfadden_conditional_1973}, or dynamic split ratio solvers such as the one presented in \citet{wright_dynamic-system-based_2018}.

\begin{figure}[htb]
\centering
\includegraphics[width=2in]{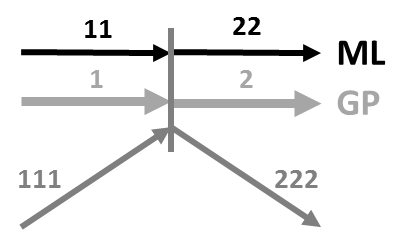}
\caption{A node where some of the input links form travel facilities with some
of the output links. ML = Managed Lane.}
\label{fig-node-1}
\end{figure}

When applied to lane choice (e.g., as in~\citet{farhi_logit_2013} or \citet{shiomi_multilane_2015}), logit models produce a set of portions in $[0,1]$, one for each lane, that sum to one.
Each lane's value is the equilibrium portion of vehicles that will travel on that lane.
In a dynamic simulation context, such as considered in this article, the differences between the logit model's equilibrium portions and the current distribution of vehicles across lanes at time $t$ are used to select split ratios at time $t$ such that the actual distribution approaches the logit equilibrium.
We argue, though, that use of this sort of split ratio solver in unmodified form might be inappropriate for computing gate split ratios in the gated-access managed lane configuration.
Unmodified, a value function for either the GP or managed lane link might include terms such as the link's speed of traffic, density, etc.
However, often a gated-access managed lane is separated from the GP lanes by some buffer zone or visibility-obstructing barrier that makes switching between the two links more hazardous than switching between two contiguous lanes.
Therefore, if one uses a logit-based model, it would be appropriate to modify the logit model's value function such that staying in the current link (e.g., the movements (1,2) and (11,22) in Figure~\ref{fig-node-1}) has some positive value for drivers, and the gains (e.g., in travel time) for ingress/egress movements (e.g., (1,22) and (11,2) in Figure~\ref{fig-node-1}) must be of more value than the staying-in-the-lane value.
We refer to this model as the \emph{inertia effect}.

We may also incorporate the inertia effect into dynamic split ratio solvers, such as the one introduced in \citet{wright_dynamic-system-based_2018}.
That dynamic split ratio solver is reviewed in Appendix \ref{app_splitratiosolver}.
At time $t$, this particular algorithm selects split ratios in an attempt to balance the density ratio at the next timestep $t+1$, $\sum_c n_l^c(t+1)/n_l^J$, where $n_l^J$ is the jam density, or the maximum number of vehicles that link $l$ can hold.
For example, if applied to the node in Figure \ref{fig-node-1}, this algorithm would attempt to make $\sum_c n_2^c(t+1)/n_l^J$ and $\sum_c n_{22}^c(t+1)/n_l^J$ as equal as possible.

Here, we modify several steps of the solver so that this equality-seeking goal is balanced with a goal towards enforcing the inertia effect.
We illustrate these changes with the particular example of the node in Figure~\ref{fig-node-1}.
Ensuring that the split ratio assignment algorithm
gives preferences to movements (1,2) and (11,22) over
(1,22) and (11,2)
can be done in step~5 of the original algorithm, setting of oriented priorities.
Specifically, we modify~\eqref{eq_oriented_priorities_undefined_sr2}.
For this particular example, the original formula gives us:
\begin{linenomath}
\begin{equation*}
\begin{array}{ll}
\gamma_{1,2}^c(k) = \tbeta_{1,2}^c(k) + \frac{\obeta_1^c(k)}{2} &
\gamma_{1,22}^c(k) = \tbeta_{1,22}^c(k) + \frac{\obeta_1^c(k)}{2}; \\
\gamma_{11,2}^c(k) = \tbeta_{11,2}^c(k) + \frac{\obeta_{11}^c(k)}{2} &
\gamma_{11,22}^c(k) = \tbeta_{11,22}^c(k) + \frac{\obeta_{11}^c(k)}{2}.
\end{array}
\end{equation*}
\end{linenomath}
Since for $k=0$ $\tbeta_{i,j}^c(0)=0$, $i=1,11$, $j=2,22$,
we get $\gamma_{1,2}^c(0) = \gamma_{1,22}^c(0) = \frac{\obeta_1^c(k)}{2}$
and $\gamma_{11,2}^c(0) = \gamma_{11,22}^c(0) = \frac{\obeta_{11}^c(k)}{2}$,
which, according to \eqref{eq_oriented_priorities_undefined_sr1},
yields $\pt_{1,2}(0)=\pt_{1,22}(0)$ and $\pt_{11,2}(0)=\pt_{11,22}(0)$.

For each input link $i$ that forms one lane with an output link $\hat{\jmath}$
and class $c$, such that $\hat{\jmath}\in V_i^c$, \eqref{eq_oriented_priorities_undefined_sr2} can be modified
as follows:
\begin{linenomath}
\begin{equation}
\gamma_{ij}^c(k) = \left\{\begin{array}{ll}
\beta_{ij}^c, & \mbox{ if split ratio is defined a priori: }
\{i,j,c\}\in\BB, \\
\tbeta_{ij}^c(k) + \obeta_i^c(k)\lambda_i^c, & 
\mbox{ $i$ and $j$ form one lane: $j=\hat{\jmath}$}, \\
\tbeta_{ij}^c(k) + \obeta_i^c(k)\frac{1-\lambda_i^c}{|V_i^c|-1}, &
\mbox{ $i$ and $j$ are in different lanes: $j\neq\hat{\jmath}$},
\end{array}\right.
\label{eq_oriented_priorities_undefined_sr22}
\end{equation}
\end{linenomath}
where the parameter $\lambda_i^c\in \left[\frac{1}{|V_i^c|}, 1\right]$ is
called the \emph{inertia coefficient}, and indicates how strong the
inertia effect is.
With $\lambda_i^c = \frac{1}{|V_i^c|}$,
\eqref{eq_oriented_priorities_undefined_sr22} reduces
to the original formula, \eqref{eq_oriented_priorities_undefined_sr2}.
With $\lambda_i^c = 1$, all the \emph{a priori} unassigned traffic from link
$i$ must stay in its lane --- be directed to output link $\hat{\jmath}$.
The choice of $\lambda_i$ lies with the modeler.
In the case of example from Figure~\ref{fig-node-1}, the modified
formula~\eqref{eq_oriented_priorities_undefined_sr22} yields:
\begin{linenomath}
\begin{equation}
\begin{array}{ll}
\gamma_{1,2}^c(k) = \tbeta_{1,2}^c(k) + \obeta_1^c(k)\lambda_1^c &
\gamma_{1,22}^c(k) = \tbeta_{1,22}^c(k) + \obeta_1^c(k)(1-\lambda_1^c); \\
\gamma_{11,2}^c(k) = \tbeta_{11,2}^c(k) + \obeta_{11}^c(k)(1-\lambda_{11}^c) &
\gamma_{11,22}^c(k) = \tbeta_{11,22}^c(k) + \obeta_{11}^c(k)\lambda_{11}^c,
\end{array}
\end{equation}
\end{linenomath}
where $\lambda_1,\lambda_{11}\in\left[\frac{1}{2}, 1\right]$,
and picking $\lambda_1>\frac{1}{2}$ ($\lambda_{11}>\frac{1}{2}$)
would give preference to movement (1,2) over (1,22)
(and (11,22) over (11,2)).

The way of choosing $\lambda_i^c$ for multiple input links is not obvious
and an arbitrary choice may result in an unbalanced flow distribution
among output links.
Therefore, we suggest picking just one input-output pair
$(\hat{\imath},\hat{\jmath})$, and for that input link setting
$\lambda_{\hat{\imath}}^c=1$,
while for other input links $i$ setting
$\lambda_i^c = \frac{1}{|V_i^c|}$, $c=1,\dots,C$.
The input link $\hat{\imath}$ must be from the lane that is expected
to have a positive net inflow of vehicles as a result of the split ratio assignment and flows computed by the node model.
So,
\begin{linenomath}
\begin{equation}
\hat{\imath} = \arg\min_{i\in \hat{U}} = 
\frac{\sum_{c: \{i,j,c\}\in\OBB}\So_i^c +
\sum_{i: \{i,j,c\}\in\BB}\sum_{c: \{i,j,c\}\in\BB} \beta_{ij}^c S_i^c}{R_j},
\label{eq_lucky_input_link}
\end{equation}
\end{linenomath}
where 
\begin{linenomath}
\begin{equation}
\hat{U} = \left\{\mbox{input links } i:\;
\exists j, c, \mbox{ s.t. } j\in V_i^c
\mbox{ and the pair of links $(i,j)$ belongs to the same lane}\right\},
\label{eq_common_lane_input_set}
\end{equation}
\end{linenomath}
and $j$ denotes the output link that is in the same lane as input link $i$.

For the particular example node in Figure~\ref{fig-node-1},
we need to determine whether flow from link 1 will proceed to link 2
or flow from link 11 to link 22, while
other \emph{a priori} undefined split ratios will be computed according to the
split ratio assignment algorithm.
If $\hat{\imath}=11$, then $\lambda_{11}=1$, $\lambda_1=\frac{1}{2}$, and
\emph{a priori} unassigned traffic in the managed lane
will stay in the managed lane ($\beta_{11,2}^c=0$), while
the \emph{a priori} unassigned traffic coming from
links~1 and~111, will be distributed between links~2 and~22
according to the dynamic split ratio solver.

On the other hand,
if $\hat{\imath}=1$, then $\lambda_{1}=1$, $\lambda_{11}=\frac{1}{2}$, and
the \emph{a priori} unassigned traffic in the GP lane
will stay in the GP lane ($\beta_{1,22}^c=0$), while
the \emph{a priori} unassigned traffic coming from
links~11 and~111, will be distributed between links~2 and~22
according to the dynamic split ratio solver.

\subsection{Smoothing Effect}
\label{sec:smoothing}
The smoothing effect is the name given to an empirically-observed phenomenon where the presence of a managed lane leads to increased capacity at bottlenecks \citep{menendez_effects_2007}.

\citet{cassidy_smoothing_2010} outline the mechanics behind the smoothing effect as follows.
They observe two sites in California where freeway bottlenecks regularly occur at peak traffic (at one site, a merge with an onramp and at the other, a curved section in the road).
At both sites, a managed lane exists; specifically the leftmost lane restricted to HOVs only during rush hour.
They find that when the HOV lane policy is enforced, the discharge of the managed lanes at the bottleneck is increased.
Specifically, they report that at the two lanes closest to the HOV lane have discharge flow increases of roughly 10-20\% \citep[Tables 1 and 2]{cassidy_smoothing_2010}.

\citet{cassidy_smoothing_2010} suggest that this increase in discharge flow is due to reduced lane changing induced by the presence of the managed lane.
To summarize their argument, restricting the innermost lane - the furthest from on- and off-ramps and usually the one with the fastest traffic speed - to certain classes of vehicles means that fewer drivers will attempt lane changes to enter that fastest lane.
This reduction in lane-changing is likely to be particularly pronounced just downstream of onramps, since normally, some drivers that enter the freeway will want to enter the fastest lane as soon as possible, which requires multiple lane changes.
Lane change maneuvers are known to have a cost to freeway performance, since a) drivers performing a lane change are effectively taking up two lanes for at least some amount of time, and b) the surrounding traffic must slow down to accommodate a lane change, potentially causing a stop-and-go wave \citep{jin_kinematic_2010}.

A recent study \citep{kim_safety_2018} has also found that on a freeway with a gated-access managed lane, onramp/offramp locations adjacent to a gate counterintuitively have a lower rate of accidents than those with no nearby gate.
This result may also be related to the presence of a managed lane somehow reducing the number of unnecessary lane changes.

We propose that the smoothing effect can be elegantly captured within the node model of a macroscopic simulation (recall from Section \ref{sec_lit_review} that the node model refers to the mathematical rule used to compute the node throughflows $f_{i,j}^c$ given the link demands and supplies at a junction).
In particular, we recall the so-called ``general class of node models (GCNM)'' introduced by \citet{tampere_generic_2011}, that serves as the base for most modern node models.
In the present paper, we will not review the GCNM node models or their derivatives (e.g., \citet{smits_family_2015,jabari_node_2016,wright_node_2017}, and others) in depth, but will discuss a particular component of this body of work that neatly captures the same phenomenon described by the smoothing effect.

Our discussion of node models so far in this paper has referred to ``supply'' as a property of links, that defines how many vehicles that link can accept as a function of its current occupancy and the link model.
Defining what occurs in situations where there is insufficient supply to accept all upstream demand is a key component of node models, and shortcomings of previous node models in this area was a key point of contrast in the original presentation of the GCNM \citep{tampere_generic_2011}.
The scheme for distribution of scarce supply in the node model, and re-distribution of ``leftover'' supply if some input link is not able to use all of its initially-apportioned supply, was termed the ``supply constraint interaction rule'' (SCIR) in \citet{tampere_generic_2011}, and one area in which node models in the GCNM family vary is the choice of SCIR \citep{smits_family_2015,wright_node_2017}.

In prior work, we have discussed a particular SCIR choice and node model constraint variously referred to as the ``conservation of turning fractions'' (CTF) or ``first-in-first-out'' (FIFO) constraint \citep{wright_dynamic_2016, wright_node_2017}.
As the name suggests, a node model that includes the FIFO constraint will require that, in congested (supply-constrained) nodes, total input-link queueing occurs when any one of an input links' destination links runs out of supply.
As an example to illustrate this idea, a FIFO constraint would mandate queueing in a GP link, if the GP link is not able to send to an adjacent managed lane link due to supply constraints in the managed lane link.
Some recent papers \citep{wright_dynamic_2016, wright_node_2017} discuss a relaxation of the FIFO constraint that we call the ``partial FIFO'' constraint.
It turns out that a node model that encodes the partial FIFO constraint is also one that, in the case of a managed lane-freeway network, can lead to the emergence of the smoothing effect.
We will illustrate this through the following example.

\subsubsection{Emergence of the smoothing effect under partial FIFO constraints}
\label{sec_smoothing_emergence}

\begin{figure}[htb]
    \centering
    \includegraphics[width=2in]{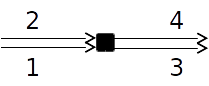}
    \caption{A simple merge-diverge node representing a GP lane and managed lane, with the opportunity for vehicles to change between the lanes.}
    \label{fig_2x2node}
\end{figure}

Consider the example node in Figure \ref{fig_2x2node}.
Here, we depict a simple managed lane-GP lane interface (with no onramps or offramps).
This may represent any node in a continuous-access managed lane, or a gate node in a gated-access managed lane.
For clarity, we will say that the links 1 and 3 represent the GP lane(s), and links 2 and 4 represent the managed lane(s).

As we have noted, lane changes are deleterious to road throughput, with several authors arguing they have a negative effect on \emph{capacity} specifically (e.g., \citet{cassidy_smoothing_2010,jin_kinematic_2010}).
This effect, of an input link having its capacity lowered via spillback in congestion, and the degree to which the capacity drop being dependent on the amount of lane-changing, can be captured in a node model of Figure \ref{fig_2x2node}'s node that incorporates the partial FIFO constraint.
In this paper, we will not go into the technical details of the partial FIFO constraint formulation, but will instead present a schematic example of how the smoothing effect emerges.
For more specific implementation details, see \citet{wright_node_2017}.

\begin{table}[h]
    \caption{Parameters and example resulting flows for situations of heavy and no lane-changing for the example node discussed in Section \ref{sec_smoothing_emergence}}
    \begin{center}
    \label{tab:params}
    \begin{tabular}{l l l l l}
    \toprule
     \multirow{2}{*}{Link Information} & $S_1 = 6000$ & $R_3 = 6000$ & &\\
     & $S_2 = 1500$ & $R_4 = 2000$ \\
     \midrule
     \rule{0pt}{4ex}
     & \multicolumn{2}{c}{Split Ratios} & \multicolumn{2}{c}{Resulting Flows} \\
        \cmidrule(lr){2-3} \cmidrule(lr){4-5}
     \multirow{3}{1.3in}{Heavy-lane-changing situation} & $\beta_{1,3} = \sfrac{2}{3}$ & $\beta_{2,3} = 0$ & $f_{1,3} = 4500$ & $f_{2,3} = 0$ \\
     & $\beta_{1,4} = \sfrac{1}{3}$ & $\beta_{2,4} = 1$ & $f_{1,4} = 500$ & $f_{2,4} = 1500$ \\
     \cmidrule(){4-5}
     & & & \multicolumn{2}{c}{$\sum_{i,j} f_{i,j} = 6500$} \\
     \rule{0pt}{4ex}
     \multirow{3}{1.3in}{No-lane-changing situation} & $\beta_{1,3} = 1$ & $\beta_{2,3} = 0$ & $f_{1,3} = 6000$ & $f_{2,3} = 0$ \\
     & $\beta_{1,4} = 0$ & $\beta_{2,4} = 1$ & $f_{1,4} = 0$ & $f_{2,4} = 1500$ \\
     \cmidrule(){4-5}
     & & & \multicolumn{2}{c}{$\sum_{i,j} f_{i,j} = 7500$} \\
     \bottomrule
    \end{tabular}
    \end{center}
\end{table}

Table 1 presents some specific numerical values we will use in our example.
In this example, we will consider only one vehicle class $c$, and so we omit the vehicle class index superscript.
Suppose the GP lane link (links 1 and 3) has three lanes, and the managed lane link (links 2 and 4) has one lane.
This is reflected in our values for the supplies and demands, with $R_3$ being three times that of $R_4$ and $S_1$ being four times $S_2$.
We use a per-lane supply of 2000, which is a common nominal value for freeway capacity in units of vehicles per hour per lane.

Table 1 also describes two situations of lane-changing (characterized by split ratios), and the resulting flows for each situation.
For the purpose of illustration (and to avoid getting bogged down in the details of the node model), we select split ratios to describe two extremes of lane-changing.
In the heavy-lane-changing situation, we are supposing that one-third of the GP-lane vehicles want to enter the leftmost lane (i.e., the managed lane).
This could be equivalent to saying that each and every vehicle in the leftmost GP lane is trying to enter the managed lane.
In the no-lane-changing situation, we are supposing that all GP vehicles and all HOV vehicles remain in their original lane.

One item that must be clarified before we can explain how we arrive at the resulting flows for these two situations is our node model's method for portioning the downstream supply among the input links competing for it.
A node model's portioning scheme makes up part of its SCIR \citep{tampere_generic_2011}.
In this example, we are trying not to get overwhelmed with the details of the node model, so we choose a very simple (and not terribly realistic) portioning scheme, where the HOV lane link is able to claim all the downstream supply it wants \emph{first}, and then the GP lane link is only able to make use of the leftover supply (c.f. the example in \cite[Section 4.2]{wright_node_2017}).
In a more realistic node model, the two input links' flows would be competing for the downstream supply, but again, our intent here is illustration of the emergence of the smoothing effect.

Consider the heavy-lane-changing situation.
To restate what is modeled by the split ratios, all the vehicles in the leftmost GP lane are trying to enter the downstream managed lane, while all the vehicles in the right two lanes are staying in the GP lanes.
However, it is clear that between the vehicles trying to change lanes into the managed lane, and the vehicles already in the managed lane, the total demand exceeds the managed lane's supply.
This means that the flow from the GP lane to the managed lane, $f_{1,3}$, is supply-constrained and not able to send all of its demand.
The implications of this supply constraint via the partial FIFO mechanism is explained next.

\begin{figure}
    \centering
    \includegraphics[width=5in]{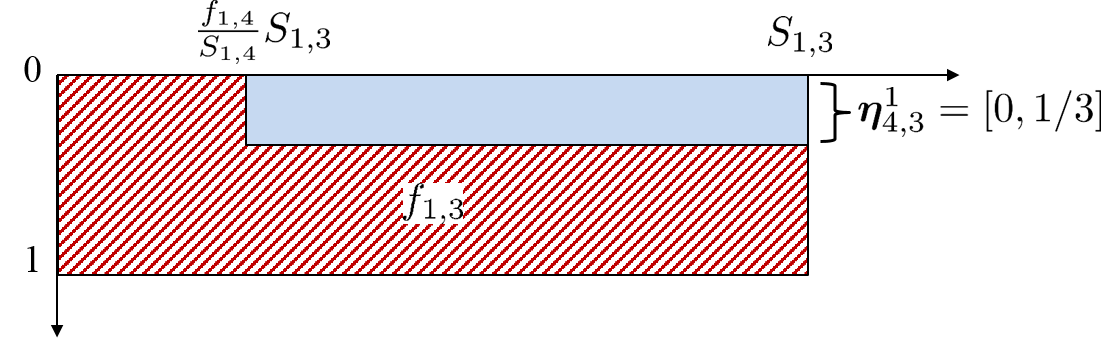}
    \caption{Diagram illustrating how the partial FIFO construction reduces the discharge flow of the GP lane under heavy lane changing in Section \ref{sec_smoothing_emergence}'s example.}
    \label{fig_restriction_smoothing}
\end{figure}

Figure \ref{fig_restriction_smoothing} presents an illustration used to explain the mechanics of the partial FIFO constraint.
This type of diagram was introduced in the original presentation of the partial FIFO constraint \citep[Section 3]{wright_node_2017}.
In this figure, $\etab_{4,3}^1$ denotes the so-called ``restriction interval,'' that describes the portion of the flow $f_{1,3}$ that becomes blocked (stuck in a queue) when the output link 4 runs out of supply.
The value of $\etab_{4,3}^1 = [0, 1/3]$ means that the leftmost third (i.e., the left lane of the three) of the GP lanes becomes queued when the managed lane runs out of downstream supply.
This value is defined by noting that only this leftmost lane serves the lane-changing movement.

The area of the hashed region labeled $f_{1,3}$ defines that flow.
The solid region defines the portion of the demand-limited flow $S_{1,3}$ that is lost due to the partial FIFO constraint and congestion in link 4 \footnote{
    To be precise, the area of the total rectangle, $S_{1,3}$, is clearly less than the maximum possible flow for $f_{1,3}$ due to supply constraint $R_3 < S_{1,3}$.
    The illustration being discussed only covers how much of the original supply becomes stuck in its origin link.
}.
The vertical length of the solid region is defined by $\etab_{4,3}^1$.
The horizontal length of the solid region has a more complex definition, and is dependent on the node's supplies, demands, and the node model's supply-portioning scheme (see \citet{wright_node_2017} for full details).
What is important in the present example is that the partial FIFO constraint says that the inability of the downstream managed lane link (link 4) to accept all the vehicles that the upstream GP lane link (link 1) wants to send it, adversely affects the discharge rate of the flow that is staying on the GP lanes.
In particular, if we use our earlier-described simple supply-portioning scheme for link 4's supply (the GP lane vehicles trying to change lanes only get the leftover supply after the upstream managed lane link has used all it wants), then we can find the leftmost boundary of the solid region as
\begin{linenomath}
\begin{equation*}
    \frac{f_{1,4}}{S_{1,4}} S_{1,3} = \frac{500}{2000}6000 = 1000
\end{equation*}
\end{linenomath}
which means that the total area of the hashed region, and the GP lane throughflow is
\begin{linenomath}
\begin{equation*}
    S_{1,3} (1) - \left( S_{1,3} - \frac{f_{1,4}}{S_{1,4}} S_{1,3} \right) | \etab_{4,3}^1 |
    = 6000 - \left(6000 - \frac{500}{2000}6000 \right) \left( \frac{1}{3} \right) = 4500.
\end{equation*}
\end{linenomath}
In other words, the heavy lane changing activated the partial FIFO constraint, which led to spillback into the GP lanes.

We can see the emergence of the smoothing effect by comparing the above heavy-lane-changing situation with the no-lane-changing situation also outlined in Table \ref{tab:params}.
This situation is much simpler to calculate the resulting flows, since both the GP lane vehicles and the managed lane vehicles wholly want to stay in their original link.
In this situation, the partial FIFO constraint does not play any role, and both input links are able to satisfy the entirety of their demands.
This leads to increased throughflows, both on the GP lanes and the node as a whole.

The dramatic increase in GP lane throughflow (1500 vehicles, or one-and-one-third of the original value of 4500) is much more than the 10-20\% increase due to the smoothing effect reported by, e.g., \citet{cassidy_smoothing_2010}.
This, though, is partially artifacts of the simplicity of our example.
Our two situations had extreme differences in lane-changing split ratios, and, as also mentioned, our supply portioning method (the GP lanes only get any leftover supply in the managed lane link) were both unrealistic but useful to demonstrate that activation of a partial FIFO constraint can be modeled wholly due to lane changing.
A more realistic model of either should result in more muted throughflow increases.

To summarize, the partial FIFO construction can describe a bottleneck that is created by heavy lane changing.
This, of course, means that a node model with a partial FIFO construction can be used to construct a smoothing effect in macroscopic simulation if lane changes between the GP lane link and the managed lane link are reduced.
To the best of our knowledge, this is the first proposed model of the smoothing effect in managed lane-freeway network simulation.
The fact that the smoothing effect emerges from a previously-introduced node model is in itself also independently interesting.
Future work should focus on more realistic calibration of the partial FIFO construction to empirically-observed patterns in the smoothing effect.

\subsubsection{Comparison with related work}
Other authors (e.g., \citet{jin_kinematic_2010,jin_multi-commodity_2013}) have considered the problem of modeling lane-changing traffic flow.
These other approaches consider how the continuum flow model (i.e., the fundamental diagram) is affected by lane-changing and weaving.
In contrast, our approach described here isolates the discharge flow drop due to lane-changing into the node model.
Appealingly, our construction does not mandate a certain link model or the entirety of the node model (recall our discussion that the SCIR is still somewhat a free parameter), which fits well into the macroscopic simulation paradigm.
The analysis of discretized forms of the continuum-theory lane-changing-flow models, and comparison with the already-discretized form presented here (as well as the reverse: this link-and-node construction here at a continuous limit) is an interesting area for potential future investigation. 

\section{Conclusion}\label{sec_conclusion}
In this paper we discussed modeling procedures for
two managed lane configurations: (1) full access,
where special traffic can switch between the GP and the managed lanes at
any node; and (2) separated, where special traffic can switch between the two
lanes only at specific nodes, called gates.
We have introduced models for the
\emph{friction effect} (Section~\ref{subsec_friction}), the
\emph{inertia effect} (Section~\ref{subsec_inertia}), and the \emph{smoothing effect} (Section \ref{sec:smoothing}).
The friction effect reflects the empirically-observed drivers' fear of moving
fast in the managed lane while traffic in the adjacent GP links moves slowly due
to congestion.
We propose modeling this by making the fundamental diagram parameters of the HOV lanes functions of the GP lanes' traffic state.
The inertia effect reflects drivers' inclination to stay in
their lane as long as possible and switch only if this would obviously
improve their travel condition.
We give an example of how to adjust a driver choice model to account for this effect.
The smoothing effect is the name for the fact that implementing a managed lane policy leads to less lane-changing, which in turn increases road capacity.
We showed that such an effect of less lane-changing leading to greater throughflows emerges in a class of ``partial FIFO'' node models.

Models of freeways with managed lanes feature many more parameters than ones with GP lanes, and calibrating them can be difficult.
The emergent phenomenological effects create complex feedback loops such that individual parameters cannot be tuned in isolation.
In the sequel \citep{wright_macroscopic_2019_pt2}, we present an iterative learning approach for estimating some of the harder-to-estimate parameters.

\section*{Acknowledgements}\label{sec_acknowledgement}
We would like to express great appreciation to several of our colleagues.
To Elena Dorogush and Ajith Muralidharan for sharing ideas, and to
Gabriel Gomes and Pravin Varaiya for their
critical reading and their help in clarifying some theoretical issues.
We would also like to thank an anonymous reviewer for suggesting we expand
our discussion of the smoothing effect.

This research was supported by the California Department of Transportation.
Previous versions of portions of this work previously appeared in the technical report \citet{horowitz_modeling_2016}.

\appendix
\section{A Link Model}\label{app_linkmodel}
\begin{figure}[ht]
\centering
\includegraphics[width=3in]{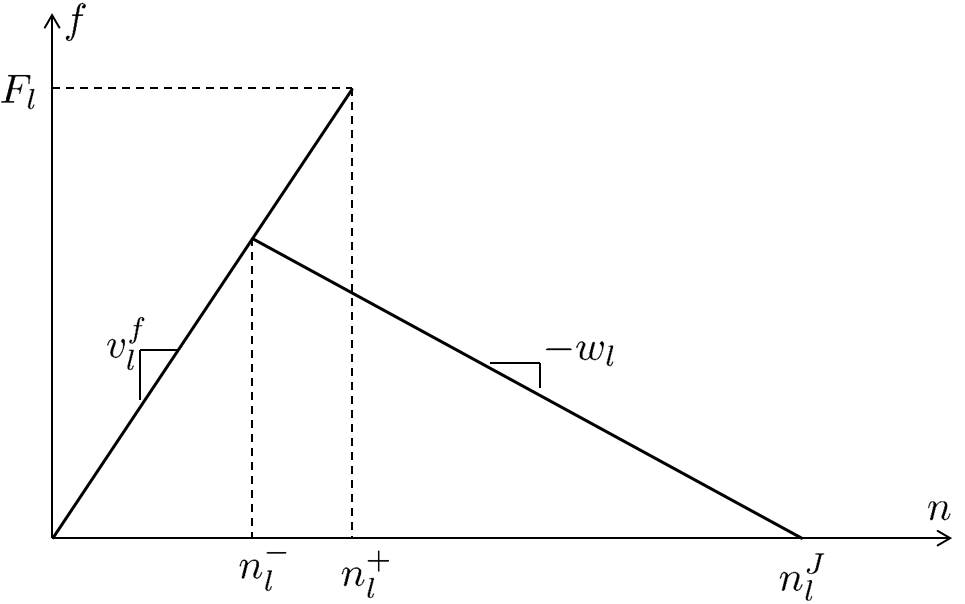}
\caption{The ``backwards lambda'' fundamental diagram.}
\label{fig-fd}
\end{figure}
For the majority of this paper, we remain agnostic as to the particular functional relationship between density $n_l$, demand $S_l$ (per commodity, $n_l^c$, $S_l^c$) and supply $R_l$, and flow $f_l$ (also called the fundamental diagram) used in our first-order macroscopic model \eqref{eq:ctmbasic}.
Where a particular fundamental diagram is required, i.e. for
the example implementation of the friction effect in \eqref{eq_friction_free_flow_speed}, \eqref{eq_friction_capacity}, \eqref{eq_friction_send_function}, and \eqref{eq_friction_threshold}, we use a fundamental diagram from~\citet{horowitz_modeling_2016},
shown in Figure~\ref{fig-fd}.
This fundamental diagram captures the traffic hysteresis behavior with the ``backwards lambda'' shape often observed in detector data \citep{koshi_findings_1983}:
\begin{linenomath}
\begin{align}
S_l^c(t) =& v_l^f(t) n_l^c(t) \min \left\{1, \frac{F_l(t)}{v_l^f(t) \sum_{c=1}^C n_l^c(t)} \right\}, \;\;\;
S_l(t) = \sum_{c=1}^C S_l^c(t),
\label{eq_lnctm_send_function} \\
R_l(t) =& \left( 1 - \theta_l(t) \right) F_l(t) + \theta_l(t) w_l(t)
  \left( n_l^J(t) - \sum_{c=1}^C n_l^c(t) \right), \label{eq_lnctm_receive_function}
\end{align}
\end{linenomath}
where, for link $l$, $F_l$ is the capacity, $v_l^f$ is the free flow speed, $w_l$ is the congestion wave speed, $n_l^J$ is the jam density, and
$n_l^{-}=\frac{w_ln_l^J}{v_l^{f} + w_l}$ and $n_l^{+}=\frac{F_l}{v_l^{f}}$ are called the \emph{low} and \emph{high critical densities}, respectively.
As written here and used in this paper, $F_l$, $v_l^f$, and $w_l$ are in units per simulation timestep.
The variable $\theta_l(t)$ is a congestion metastate of $l$, which encodes the hysteresis:
\begin{linenomath}
\begin{equation}
\theta_l(t) = 
\begin{cases}
0 & n_l(t)\leq n^-_l,\\
1 & n_l(t)> n^+_l,\\
\theta_l(t-1) & n_l^-<n_l(t)\leq n_l^+,
\end{cases}
\label{eq_metastate}
\end{equation}
\end{linenomath}
where $n_l(t) = \sum_{c=1}^C n_l^c(t)$.

Examining~\eqref{eq_metastate} and~\eqref{eq_lnctm_receive_function}, we see that when a link's density goes above $n_l^{+}$ (i.e., when it becomes congested), its ability to receive flow is reduced until the density falls below $n_l^{-}$.

An image of~\eqref{eq_lnctm_send_function}
and~\eqref{eq_lnctm_receive_function} overlaid on each other,
giving a schematic image of the fundamental diagram,
is shown in Figure~\ref{fig-fd}.
Unless $n_l^{-}=n_l^{+}$, when it assumes triangular shape,
the fundamental diagram is not a function of density alone (i.e., without $\theta_l(t)$):
$n_l(t)\in\left(n_l^{-}, n_l^{+}\right]$ admits two possible flow values.

\section{Dynamic Split Ratio Solver}\label{app_splitratiosolver}
Throughout this article, we have made reference to a dynamic-system-based method for solving for partially- or fully-undefined split ratios from \citet{wright_node_2017}.
This split ratio solver is designed to implicitly solve the logit-based split ratio problem
\begin{equation}
  \beta_{ij}^c = \frac{\exp \left( \frac{\sum_{i=1}^M \sum_{c=1}^C S_{ij}^c}{R_j} \right)}{
  \sum_{j'=1}^N \exp \left( \frac{\sum_{i=1}^M \sum_{c=1}^C S_{ij'}^c}{R_{j'}} \right)}, \label{eq:splitratio_demandsupply}
\end{equation}
which cannot be solved explicitly, as the $S_{ij}^c$'s are also functions of the $\beta_{ij}^c$'s.
The problem \eqref{eq:splitratio_demandsupply} is chosen to be a node-local problem that does not rely on information from the link model (beyond supplies and demands), and is thus independent of the choice of link model \citep{wright_node_2017}.

The solution algorithm is as follows, reproduced from \citet{wright_node_2017}. More discussion is available in the reference.

\begin{itemize}
\item Define the set of commodity movements for which split ratios
are known as $\BB=\left\{\left\{i,j,c\right\}: \; \beta_{ij}^c \in [0,1]\right\}$,
and the set of commodity movements for which split ratios are to be
computed as $\OBB=\left\{\left\{i,j,c\right\}: \; \beta_{ij}^c \mbox{ are unknown}\right\}$.

\item For a given input link $i$ and commodity $c$ such that $S_i^c=0$,
assume that all split ratios are known: $\{i,j,c\}\in\BB$.\footnote{If split
ratios were undefined in this case, they could be assigned arbitrarily.}

\item Define the set of output links for which there exist unknown
split ratios as $V=\left\{j: \; \exists \left\{i,j,c\right\}\in\OBB\right\}$.

\item Assuming that for a given input link $i$ and commodity $c$,
the split ratios must sum up to 1, define the unassigned portion
of flow by $\obeta_i^c=1-\sum_{j:\{i,j,c\}\in\BB}\beta_{ij}^c$.

\item For a given input link $i$ and commodity $c$ such that there exists
at least one commodity movement $\{i,j,c\}\in\OBB$, assume $\obeta_i^c>0$, otherwise the
undefined split ratios can be trivially set to 0.

\item For every output link $j\in V$, define the set of input links
that have an unassigned demand portion directed toward this output link
by $U_j=\left\{i: \; \exists\left\{i,j,c\right\}\in\OBB\right\}$.

\item For a given input link $i$ and commodity $c$, define the set
of output links for which split ratios for which are to be computed as
$V_i^c = \left\{j: \; \exists i\in U_j\right\}$,
and assume that if nonempty, this set contains at least two elements,
otherwise a single split ratio can be trivially set equal to $\obeta_i^c$.

\item Assume that input link priorities are non-negative, $p_i\geq 0$,
	$i=1,\dots,M$, and $\sum_{i=1}^M p_i = 1$.

\item Define the set of input links with zero priority:
$U_{zp} = \left\{i:\;p_i=0\right\}$.
To enable split ratio assignment for inputs with zero priorities,
perform regularization:
\be
\pt_i = p_i\left(1-\frac{|U_{zp}|}{M}\right)+ \frac{1}{M}\frac{|U_{zp}|}{M}=
p_i\frac{M-|U_{zp}|}{M} + \frac{|U_{zp}|}{M^2},
\label{priority_regularization}
\ee
where $|U_{zp}|$ denotes the number of elements in set $U_{zp}$.
Expression~\eqref{priority_regularization} implies that the regularized
input priority $\pt_i$ consists of two parts:
(1) the original input priority $p_i$ normalized to the portion of 
input links with positive priorities; and
(2) uniform distribution among $M$ input links, $\frac{1}{M}$, 
normalized to the portion of input links with zero priorities.

Note that the regularized priorities $\pt_i> 0$, $i=1,\dots,M$, and $\sum_{i=1}^M \pt_i = 1$.
\end{itemize}

The algorithm for distributing $\obeta_i^c$ among the commodity movements
in $\OBB$ (that is, assigning values to the a priori unknown split ratios)
aims at maintaining output links as uniform in their demand-supply ratios as possible.
At each iteration $k$, two quantities are identified: $\mu^+(k)$, which is the largest \emph{oriented} demand-supply ratio produced by the split ratios that have been assigned so far, and $\mu^-(k)$, which is the smallest oriented demand-supply ratio whose input link, denoted $i^-$, still has some unclaimed split ratio.
Once these two quantities are found, the commodity $c^-$ in $i^-$ with the smallest unallocated demand has some of its demand directed to the $j$ corresponding to $\mu^-(k)$ to bring $\mu^-(k)$ up to $\mu^+(k)$ (or, if this is not possible due to insufficient demand, all such demand is directed).

To summarize, in each iteration $k$, the algorithm attempts to bring the smallest oriented demand-supply ratio $\mu^+(k)$ up to the largest oriented demand-supply ratio $\mu^-(k)$.
If it turns out that all such oriented demand-supply ratios become perfectly balanced, then the demand-supply ratios $(\sum_i \sum_c S_{ij}^c) / R_j$ are as well.

The algorithm is:

\begin{enumerate}
\item Initialize:
\begin{eqnarray*}
\tbeta_{ij}^c(0) & := & \left\{\begin{array}{ll}
\beta_{ij}^c, & \mbox{ if } \{i,j,c\}\in\BB,\\
0, & \mbox{ otherwise};\end{array}\right. \\
\obeta_i^c(0) & := & \obeta_i^c; \\
\Ut_j(0) & = &  U_j; \\
\Vt(0) & = & V; \\
k & := & 0,
\end{eqnarray*}
Here $\Ut_j(k)$ is the remaining set of input links with some unassigned demand,
which may be directed to output link $j$; and
$\Vt(k)$ is the remaining set of output links, to which the still-unassigned
demand may be directed.

\item If $\Vt(k)=\emptyset$, stop.
The sought-for split ratios are $\left\{\tbeta_{ij}^c(k)\right\}$,
$i=1,\dots,M$, $j=1,\dots,N$, $c=1,\dots,C$.

\item Calculate the remaining unallocated demand:
\[
\So_i^c(k) = \obeta_i^c(k) S_i^c, \;\;\; i=1,\dots,M, \;\; c=1,\dots,C.
\]

\item For all input-output link pairs,
calculate oriented demand:
\[ \St_{ij}^c(k) = \tbeta_{ij}^c(k) S_i^c. \]

\item For all input-output link pairs, calculate oriented priorities:
\begin{eqnarray}
\pt_{ij}(k) & = & \pt_i\frac{\sum_{c=1}^C \gamma_{ij}^c S_i^c}{
\sum_{c=1}^C S_i^c}
\label{eq_oriented_priorities_undefined_sr1} \\
\mbox{with} & & \nonumber \\
\gamma_{ij}^c(k) & = & \left\{\begin{array}{ll}
\beta_{ij}^c, & \mbox{ if split ratio is defined a priori: }
\{i,j,c\}\in\BB, \\
\tbeta_{ij}^c(k) + \frac{\obeta_i^c(k)}{|V_i^c|}, &
\mbox{ otherwise},\end{array}\right.
\label{eq_oriented_priorities_undefined_sr2}
\end{eqnarray}
where $|V_i^c|$ denotes the number of elements in the set $V_i^c$.
Examining the
expression~\eqref{eq_oriented_priorities_undefined_sr1}-\eqref{eq_oriented_priorities_undefined_sr2}, one can see that the
split ratios $\tbeta_{ij}^c(k)$, which are not fully defined yet,
are complemented with a fraction of $\obeta_i^c(k)$ inversely proportional
to the number of output links among which the flow of commodity $c$
from input link $i$ can be distributed.

Note that in this step we are using \emph{regularized} priorities $\pt_i$
as opposed to the original $p_i$, $i=1,\dots,M$.
This is done to ensure that inputs with $p_i=0$ are not ignored
in the split ratio assignment.

\item Find the largest oriented demand-supply ratio:
\[
\mu^+(k) = \max_{j} \max_{i}
\frac{\sum_{c=1}^C \St_{ij}^c(k)}{\pt_{ij}(k)R_j}\sum_{i\in U_j}\pt_{ij}(k).
\]

\item Define the set of all output links in $\Vt(k)$, where the minimum of
the oriented demand-supply ratio is achieved:
\[
Y(k) = \arg\min_{j\in\Vt(k)}\min_{i\in\Ut_j(k)}
\frac{\sum_{c=1}^C \St_{ij}^c(k)}{\pt_{ij}(k)R_j}
\sum_{i\in U_j}\pt_{ij}(k),
\]
and from this set pick the output link $j^-$ with the smallest
output demand-supply ratio (when there are multiple
minimizing output links, any of the minimizing output links
may be chosen as $j^-$):
\[
j^- = \arg\min_{j\in Y(k)}
\frac{\sum_{i=1}^M\sum_{c=1}^C\St_{ij}^c(k)}{R_j}.
\]

\item Define the set of all input links, where the minimum of
the oriented demand-supply ratio for the output link $j^-$ is achieved:
\[
W_{j^-}(k) = \arg\min_{i\in\Ut_{j^-}(k)}
\frac{\sum_{c=1}^C \St_{ij^-}^c(k)}{\pt_{ij^-}(k)R_{j^-}}
\sum_{i\in U_{j^-}}\pt_{ij^-}(k),
\]
and from this set pick the input link $i^-$ and commodity $c^-$
with the smallest remaining unallocated demand:
\[
\{i^-, c^-\} = \arg\min_{\begin{array}{c}
i\in W_{j^-}(k),\\
c:\obeta_{i^-}^c(k)>0\end{array}} \So_i^c(k).
\]

\item Define the smallest oriented demand-supply ratio:
\[
\mu^-(k) = 
\frac{\sum_{c=1}^C \St_{i^-j^-}^c(k)}{\pt_{i^-j^-}(k)R_{j^-}}
\sum_{i\in U_{j^-}}\pt_{ij-}(k).
\]
\begin{itemize}
\item If $\mu^-(k) = \mu^+(k)$, the oriented demands created by
the split ratios that have been assigned as of iteration $k$,
$\tbeta_{ij}^c(k)$, are perfectly balanced among the output links,
and to maintain this, all remaining unassigned split ratios should
be distributed proportionally to the allocated supply:
\begin{eqnarray}
\tbeta_{ij}^{c}(k+1) & = & \tbeta_{ij}^{c}(k) + 
\frac{R_j}{\sum_{j'\in V_{i}^{c}(k)}R_{j'}}
\obeta_{i}^{c}(k), \;\;\; 
c:~\obeta_i^c(k) > 0, \;\; i \in \Ut_j(k), \;\; j\in \Vt(k);
\label{eq_sr_assign_1}\\
\obeta_{i}^{c}(k+1) & = & 0, \;\;\; c:~\obeta_i^c(k) > 0,
\;\; i \in \Ut_j(k), \;\; j\in \Vt(k); \nonumber\\
\Ut_j(k+1) & = & \emptyset, \;\;\; j\in\Vt(k); \nonumber \\
\Vt(k+1) & = & \emptyset. \nonumber
\end{eqnarray}
If the algorithm ends up at this point, we have emptied $\Vt(k+1)$ and are done.
\item Else, assign:
\begin{eqnarray}
\Delta\tbeta_{i^-j^-}^{c^-}(k) & = & \min\left\{\obeta_{i^-}^{c^-}(k),\;\;
\left(\frac{\mu^+(k)\pt_{i^-j^-}(k)R_{j^-}}{
\So_{i^-}^{c^-}(k) \sum_{i\in U_{j^-}}\pt_{ij^-}(k)} -
\frac{\sum_{c=1}^C\St_{i^-j^-}^c(k)}{\So_{i^-}^{c^-}(k)}\right)\right\};
\label{eq_sr_assign_3}\\
\tbeta_{i^-j^-}^{c^-}(k+1) & = & \tbeta_{i^-j^-}^{c^-}(k) + 
\Delta\tbeta_{i^-j^-}^{c^-}(k);
\label{eq_sr_assign_4}\\
\obeta_{i^-}^{c^-}(k+1) & = & \obeta_{i^-}^{c^-}(k) -
\Delta\tbeta_{i^-j^-}^{c^-}(k); \label{eq_sr_assign_5} \\
\tbeta_{ij}^{c}(k+1) & = & \tbeta_{ij}^{c}(k) \mbox{ for }
\{i,j,c\}\neq\{i^-,j^-,c^-\}; \nonumber\\
\obeta_i^c(k+1) & = & \obeta_i^c(k) \mbox{ for } \{i,c\}\neq\{i^-,c^-\};
\nonumber \\
\Ut_j(k+1) & = & \Ut_j(k) \setminus \left\{i: \;
\obeta_{i}^c(k+1) = 0, \; c=1,\dots,C\right\}, \;\;\; j\in \Vt(k); \nonumber\\
\Vt(k+1) & = & \Vt(k) \setminus \left\{j: \; \Ut_j(k+1)=\emptyset\right\}.
\nonumber
\end{eqnarray}
In~\eqref{eq_sr_assign_3}, we take the minimum of the remaining unassigned
split ratio portion $\obeta_{i^-}^{c^-}(k)$ and the split ratio portion
needed to equalize $\mu^-(k)$ and $\mu^+(k)$.
To better understand the latter, the second term in $\min\{\cdot,\cdot\}$ can
be rewritten as:
\[
\frac{\mu^+(k)\pt_{i^-j^-}(k)R_{j^-}}{
\So_{i^-}^{c^-}(k) \sum_{i\in U_{j^-}}\pt_{ij^-}(k)} -
\frac{\sum_{c=1}^C\St_{i^-j^-}^c(k)}{\So_{i^-}^{c^-}(k)} =
\left(\frac{\mu^+(k)}{\mu^-(k)}-1\right)
\left(\sum_{c=1}^C\St_{i^-j^-}^c(k)\right)
\frac{1}{\So_{i^-}^{c^-}(k)}.
\]
The right hand side of the last equality can be interpreted as:
flow that must be assigned for input $i^-$, output $j^-$ and commodity $c^-$
to equalize $\mu^-(k)$ and $\mu^+(k)$ minus flow that is already assigned
for $\{i^-,j^-,c^-\}$, divided by the remaining unassigned portion
of demand of commodity $c^-$ coming from input link $i^-$.

In~\eqref{eq_sr_assign_4} and~\eqref{eq_sr_assign_5}, the assigned
split ratio portion is incremented and the unassigned split ratio portion is
decremented by the computed $\Delta\tbeta_{i^-j^-}^{c^-}(k)$.
\end{itemize}

\item Set $k := k+1$ and return to step 2.
\end{enumerate}

\bibliographystyle{abbrvnat}

\end{document}